# Dynamics of the antiferromagnetic skyrmion induced by a magnetic anisotropy gradient

Laichuan Shen [1, *], Jing Xia [2, *], Guoping Zhao [1, †], Xichao Zhang [2, 3], Motohiko Ezawa [4], Oleg A. Tretiakov [5, 6], Xiaoxi Liu [3], and Yan Zhou [2, †]

[1]*College of Physics and Electronic Engineering, Sichuan Normal University, Chengdu 610068, China*

[2]*School of Science and Engineering, The Chinese University of Hong Kong, Shenzhen, Guangdong 518172, China*

[3]*Department of Electrical and Computer Engineering, Shinshu University, 4-17-1 Wakasato, Nagano 380-8553, Japan*

[4]*Department of Applied Physics, The University of Tokyo, 7-3-1 Hongo, Tokyo 113-8656, Japan*

[5]*Institute for Materials Research, Tohoku University, Sendai 980-8577, Japan*

[6]*School of Physics, The University of New South Wales, Sydney 2052, Australia*

\* These authors contributed equally to this work.

† Authors to whom correspondence should be addressed:

E-mail (G.Z.): zhaogp@uestc.edu.cn & E-mail (Y.Z.): zhouyan@cuhk.edu.cn

**ABSTRACT**

The dynamics of antiferromagnets is a current hot topic in condensed matter physics and spintronics. However, the dynamics of insulating antiferromagnets cannot be excited by an electric current, which is a method usually used to manipulate ferromagnetic metals. Here, we propose to use the voltage-controlled magnetic anisotropy gradient as an excitation source to manipulate insulating antiferromagnetic textures. We analytically and numerically study the dynamics of an antiferromagnetic skyrmion driven by a magnetic anisotropy gradient. Our analytical calculations demonstrate that such a magnetic anisotropy gradient can effectively drive an antiferromagnetic skyrmion towards the area of lower magnetic anisotropy. The micromagnetic simulations are in good agreement with our analytical solution. Furthermore, the magnetic anisotropy gradient induced velocity of an antiferromagnetic skyrmion is compared with that of a ferromagnetic skyrmion. Our results are useful for the understanding of antiferromagnetic skyrmion dynamics and may open a new way for the design of antiferromagnetic spintronic devices.

***Introduction.*** – Skyrmions are topologically protected magnetic textures that have been experimentally observed in chiral materials (for examples, MnSi [1], Fe$_{1-x}$Co$_x$Si [2], FeGe [3], Cu$_2$OSeO$_3$ [4], and ultrathin Pt/Co/MgO nanostructures [5]) without inversion symmetry. Skyrmion-based racetrack memory [6-9] has attracted great interest, where magnetic skyrmions are used as information carriers. Compared to domain walls, skyrmions have advantages of nanoscale size and topological stability. Moreover, experiments have shown that skyrmions require a depinning current density of ~ $10^6$ A/m$^2$ [10, 11], which is much smaller than that of domain walls (~ $10^{10} - 10^{12}$ A/m$^2$) [12, 13]. Various methods have been proposed to drive magnetic skyrmions, such as by employing electric currents [7-9, 14], spin waves [15], magnetic field gradients [16], and temperature gradients [17]. Recently, it has been experimentally and theoretically demonstrated that voltage-controlled magnetic anisotropy (VCMA) gradient can be used to drive and control the motion of ferromagnetic skyrmions [18-21]. Such a VCMA effect has been experimentally proven in many systems, such as in Ir/CoFeB/MgO [22], Ta/CoFeB/MgO [23], Pt/Co [24, 25], and Fe/Co/MgO [26] systems.

On the other hand, antiferromagnets with compensated magnetic sublattices have attracted great attention [27-41] and are promising as candidate materials for advanced spintronic devices due to their zero stray fields and ultrafast magnetization dynamics. Theoretical calculations demonstrate the existence of stable skyrmions in antiferromagnets [42-44]. Compared to ferromagnetic (FM) skyrmions, antiferromagnetic (AFM) skyrmions are ideal information carriers because they have no skyrmion Hall effect [42-45]. However, it is difficult to manipulate the AFM skyrmion using the magnetic field due to the zero net magnetic moments of a perfect antiferromagnet, and also the dynamics of the insulating antiferromagnet cannot be induced by an electric current [32, 46-48]. A vital question is how to drive an insulating AFM skyrmion. Therefore, alternative methods are crucial and have been explored, for example, using temperature gradients [30, 49] and spin waves [27].

In this work, we analytically and numerically study the magnetization dynamics of an AFM skyrmion under a voltage-controlled magnetic anisotropy gradient. Such a magnetic anisotropy gradient is a new method to manipulate an AFM texture and could be used in the insulating AFM materials. Our results show that an AFM skyrmion will move towards the area with lower anisotropy, and the speed of an AFM skyrmion can reach 500 m/s driven by a magnetic anisotropy gradient in principle.

*Model and simulation.* – We consider an AFM film with two sublattices having magnetic moments $M_1(r, t)$ and $M_2(r, t)$, $|M_1| = |M_2| = M_S/2$ with the saturation magnetization $M_S$. The total magnetization is $M(r, t) = M_1(r, t) + M_2(r, t)$ and the staggered magnetization is $l(r, t) = M_1(r, t) - M_2(r, t)$, where the former is related to the canting of magnetic moments, and the latter gives the unit Néel vector $n(r, t) = l(r, t)/l$ ($l = |l(r, t)|$) that could be used to describe configurations of AFM skyrmions.

The AFM free energy can be written as [27, 32, 46, 47, 50-52]

$$E = \int dV [A_h M^2 + A(\nabla n)^2 - K n_z^2 + w_D], \tag{1}$$

where $A_h$ and $A$ stand for the homogeneous and inhomogeneous exchange constants, respectively, $K$ is the perpendicular magnetic anisotropy (PMA) constant, which changes linearly with the longitudinal spatial coordinate $x$ [see Fig. 1(a)], i.e., $K(x) = K_0 - x \cdot dK/dx$, and $w_D$ represents the energy density arising from the interfacial Dzyaloshinskii-Moriya interaction (DMI) [53-55], which stabilizes the Néel-type skyrmion [56] and is written as [51, 52, 57, 58]

$$w_D = D[n_z \nabla \cdot n - (n \cdot \nabla) n_z], \tag{2}$$

where $D$ is the DMI constant and $\mathbf{n} = \sin\theta(r)\cos\varphi \mathbf{e}_x + \sin\theta(r)\sin\varphi \mathbf{e}_y + \cos\theta(r)\mathbf{e}_z$ with the polar coordinates $(r, \varphi)$. In this paper, the magnetic anisotropy gradient is assumed to be induced by applying a voltage to the sample with a wedged insulating layer [see Fig. 1(a)] [18, 59, 60] and $dK/dx = 100$ GJ/m$^4$ unless otherwise addressed, which means that the magnetic anisotropy decreases by 10 kJ/m$^3$ for every 100 nm increase in the spatial coordinate $x$. Such an anisotropy change of 10 kJ/m$^3$ can be generated by a voltage of 0.1 V, where the experimental VCMA coefficient of 100 fJ/(V m) [22, 23] and the insulating layer thickness of ~ 1 nm are adopted.

Eq. (1) is widely used for the AFM free energy [51, 52, 57, 58], from which more specific forms can be obtained. In particular, assuming that $\mathbf{M} = 0$ and an isolated AFM skyrmion [Fig. 1(c)] is in a thin film, Eq. (1) can be simplified to a one-dimensional form, which is identical to the free energy of a FM skyrmion [Fig. 1(b)] [61]. Further, using variational calculus yields

$$\frac{d^2\theta}{dr^2} + \frac{1}{r}\frac{d\theta}{dr} = \left(\frac{1}{r^2} + \frac{K}{A}\right)\sin\theta\cos\theta - \frac{D\sin^2\theta}{Ar}, \qquad (3)$$

where $\theta$ is the angle between $\mathbf{n}$ and the $z$-axis. Solving Eq. (3), one can obtain the profile of a metastable AFM skyrmion [51].

The variational derivatives of Eq. (1) yield the effective fields $\mathbf{f}_M = -\delta E/\mu_0\delta\mathbf{M}$ and $\mathbf{f}_n = -\delta E/\mu_0\delta\mathbf{n}$ with the vacuum permeability constant $\mu_0$, which are indispensable in magnetization dynamics of the AFM system. The magnetization dynamics in ferromagnets is governed by Landau-Lifshitz-Gilbert (LLG) equation [62], whereas in AFM system, the dynamics of $\mathbf{M}$ and $\mathbf{n}$ are controlled by the following two coupled equations [27, 46],

$$\dot{\mathbf{n}} = (\gamma \mathbf{f}_M - G_1 \dot{\mathbf{M}}) \times \mathbf{n}, \qquad (4a)$$

$$\dot{\mathbf{M}} = (\gamma \mathbf{f}_n - G_2 \dot{\mathbf{n}}) \times \mathbf{n} + (\gamma \mathbf{f}_M - G_1 \dot{\mathbf{M}}) \times \mathbf{M}, \qquad (4b)$$

where $\gamma$ is the gyromagnetic ratio, and $G_1$ and $G_2$ are the damping parameters. Based on Eq. (4), one can numerically simulate the evolution of staggered magnetization in antiferromagnets and also obtain a dynamic equation expressed by the unit Néel vector $\boldsymbol{n}$ (see Ref. [63] for details) [46, 47],

$$(1 + G_1 G_2)\ddot{\boldsymbol{n}}/\gamma = \frac{2A_h}{\mu_0}(\gamma \boldsymbol{f_n} - G_2 \dot{\boldsymbol{n}}). \tag{5}$$

Our analytical solutions for an AFM skyrmion driven by a magnetic anisotropy gradient are obtained based on Eq. (5).

*Analytical steady-motion velocity of an AFM skyrmion driven by a PMA gradient.* – In the following, only a voltage-controlled PMA gradient along the *x*-direction is taken into account and other driving sources (e.g., currents) are not considered. Assuming that an AFM skyrmion is rigid and will eventually move steadily in the thin film, we take the scalar product of Eq. (5) with $\partial_i \boldsymbol{n}$ and then integrate over the space, it gives the AFM skyrmion velocity induced by a PMA gradient (see Ref. [63] for details),

$$v_{x,\text{AFM}} = \frac{\gamma u}{\mu_0 M_S \alpha d} \frac{dK}{dx}, \tag{6a}$$

$$v_{y,\text{AFM}} = 0, \tag{6b}$$

where *u* and *d* are calculated as,

$$d = \int dx dy\, \partial_x \boldsymbol{n} \cdot \partial_x \boldsymbol{n} = \int dx dy\, \partial_y \boldsymbol{n} \cdot \partial_y \boldsymbol{n}, \tag{7a}$$

$$u = \int dx dy (1 - n_z^2), \tag{7b}$$

and $\alpha = G_2/l$. Equation (6a) shows that the AFM skyrmion moves towards the area with lower magnetic anisotropy and the velocity is proportional to $1/\alpha$ and $dK/dx$. It also means that applying an opposite voltage can drive an AFM skyrmion towards the inverse direction since an opposite voltage results an inverse *dK/dx* [25].

It can be seen from Eq. (7) that the values of *u* and *d* are determined by the profile of AFM skyrmions. The profile of the skyrmion can be approximated by [61],

$$Cd\theta/dr = -\sin\theta, \tag{8}$$

Page **6** of **19**

where $C$ is a constant, which equals the domain wall width parameter $\Delta = (A/K)^{1/2}$ for the straight domain wall. Then we can find

$$u \approx 4\pi R_s C, \quad d \approx 2\pi(R_s/C + C/R_s), \tag{9}$$

where $R_s$ is the skyrmion radius and $\sin\theta \neq 0$ only for $r \approx R_s$ [61]. The skyrmion radius $R_s$ is given by [61]

$$R_s \approx \frac{\Delta}{\sqrt{2 - 2D/D_c}}, \tag{10}$$

where $D_c = 4(AK)^{1/2}/\pi$. In order to obtain the coefficient $C$, Eq. (3) is combined with Eq. (8) and integrated over one period. Then, $C$ is obtained as (see Ref. [63] for details)

$$C \approx \frac{\Delta}{\left(\Delta^2/R_s^2 + 1\right)^{1/2}}. \tag{11}$$

Substituting it into Eq. (6a), the velocity of skyrmion takes the form

$$v_{x,\text{AFM}} \approx \frac{2\gamma}{\mu_0 M_S \alpha} \frac{dK}{dx} \frac{R_s^2}{R_s^2/\Delta^2 + 2}. \tag{12}$$

Eq. (12) shows that the velocity of an AFM skyrmion also increases for larger skyrmions, which can be resulted from increasing $D$ or decreasing $K$.

*Numerical velocity of an AFM skyrmion driven by a PMA gradient.* – In order to verify the analytical results, we now simulate the motion of the Néel-type AFM skyrmion based on Eq. (4). A metastable AFM skyrmion is relaxed initially and then a PMA gradient is applied. The velocity components are $(v_x, v_y) = (\dot{R}_x, \dot{R}_y)$, where $(R_x, R_y)$ is the guiding center of skyrmion and defined as

$$R_i = \frac{\int dx dy [i \mathbf{n} \cdot (\partial_x \mathbf{n} \times \partial_y \mathbf{n})]}{\int dx dy [\mathbf{n} \cdot (\partial_x \mathbf{n} \times \partial_y \mathbf{n})]}, \quad i = x, y. \tag{13}$$

Figure 2 shows $v_x$ and $R_x$ of an AFM skyrmion as functions of time for $dK/dx = 100$ GJ/m$^4$ and $\alpha = 0.002$. In Fig. 2(a), the AFM skyrmion is first accelerated to ~ 450 m/s in 0.2 ns and then increases slowly to 504 m/s by $t = 0.5$ ns. The velocity cannot reach a constant value due to the change of the skyrmion size induced by the decreasing $K$. As shown in Fig.

2(c), the radius of the AFM skyrmion at $t = 0.5$ ns is larger than that of the initial state. When the AFM skyrmion moves in the positive $x$-direction, the decreasing $K$ results in the increase of skyrmion size, giving rise to the slow increase of the speed. Such an effect also exists in the case of FM skyrmion driven by a magnetic anisotropy gradient, as reported in Ref. [19]. Figure 2(b) shows the dependence of AFM skyrmion velocity on magnetic anisotropy $K_0$. The velocity increases with the decrease of $K_0$ since a decreasing $K_0$ results in a larger skyrmion, as shown in Figs. 2(c) and 2(d). The analytical velocities given by Eq. (12) are also shown in Fig. 2(b), where $R_s$ is adopted as the skyrmion radius at $t = 0$ ns. It can be seen that the analytical and numerical velocities are in good agreement, especially for the larger $K_0$ since the change of skyrmion size is negligible when $K_0 > 0.8$ MJ/m$^3$.

Figure 3 shows the dependences of the velocity on the anisotropy gradient and damping constant. To obtain the steady velocity, we adopt $K_0 = 0.8$ MJ/m$^3$ and $\alpha = 0.02 \sim 0.2$. The AFM skyrmion moves slowly and the change of skyrmion size is negligible. Thus, a steady velocity can be reached at $t = 0.2$ ns (see Ref. [63] for details). The velocity of an AFM skyrmion is proportional to the anisotropy gradient $dK/dx$ [see Fig. 3(a)] and $1/\alpha$ [see Fig. 3(b)], as expected from Eq. (6). It can be seen that numerical results are in a good agreement with the analytical results. When $dK/dx = 100$ GJ/m$^3$ and $\alpha = 0.1$, the velocity of an AFM skyrmion reaches 8.17 m/s. We also simulate the current-induced motion of an AFM skyrmion for the purpose of comparison. It is found that the AFM skyrmion can gain the velocity of 8.17 m/s when the current density $j = 9.1 \times 10^9$ A/m$^2$. The simulation details of current-induced motion are given in Ref. [63]. It is worth mentioning that compared to the current, the VCMA gradient as a driving source has some advantages in applications, for instance, reducing the Joule heating. The wedge heterostructures with finite length can be used to build the horizontal racetrack memory, as proposed in Ref. [64]. When these devices are chained together, the AFM or FM skyrmion will stop at the joint due to the increase of magnetic anisotropy, which can also be used to build skyrmion-based diode in principle [65].

The velocities of a FM skyrmion driven by a PMA gradient are also shown in Fig. 3. The simulation details are given in Ref. [63]. It can be seen that the velocity of AFM skyrmion is much larger than that of FM skyrmion under the same PMA gradient. For a FM skyrmion, the velocity can be obtained by solving the Thiele equation [66] (see Ref. [63] for details),

$$v_{x,\text{FM}} = \frac{\alpha d_F \gamma u_F}{\mu_0 M_S [(4\pi Q)^2 + (\alpha d_F)^2]} \frac{dK_F}{dx}, \quad v_{y,\text{FM}} = \frac{4\pi Q \gamma u_F}{\mu_0 M_S [(4\pi Q)^2 + (\alpha d_F)^2]} \frac{dK_F}{dx}, \quad (14)$$

where $Q$ is the ferromagnetic topological number, $u_F = \int dxdy(1 - m_z^2)$ and $d_F = \int dxdy\, \partial_x \boldsymbol{m} \cdot \partial_x \boldsymbol{m} = \int dxdy\, \partial_y \boldsymbol{m} \cdot \partial_y \boldsymbol{m}$. According to Eq. (6), the AFM skyrmion speed is larger than the FM skyrmion speed $\sqrt{v_{x,\text{FM}}^2 + v_{y,\text{FM}}^2} = \frac{\gamma u_F}{\mu_0 M_S} \frac{1}{\sqrt{(4\pi Q)^2 + (\alpha d_F)^2}} \frac{dK_F}{dx}$ with the same parameters. Moreover, the directions of AFM and FM skyrmion motion are significantly different. In particular, for a very small damping, a FM skyrmion moves almost along the y-direction, i.e., perpendicular to the gradient direction, similar to the case where the magnetic field gradient is adopted to drive a FM skyrmion [16, 67]. For small $\alpha$, the velocity of a FM skyrmion can be approximated as $v_{x,\text{FM}} \approx 0$, $v_{y,\text{FM}} \approx \frac{\gamma u_F}{\mu_0 M_S 4\pi Q} \frac{dK_F}{dx}$, which shows that the FM skyrmion moves perpendicular to the gradient direction. The results are consistent with those reported in Ref. [19]. Meanwhile, an AFM skyrmion moves along the x-direction, i.e., parallel to the gradient direction.

The velocities in Fig. 3 show that the AFM skyrmion has two obvious advantages compared to the FM skyrmion. First, the AFM skyrmion has no velocity in the y-direction ($v_y$ = 0) and moves perfectly parallel to the racetrack, so that it will not be destroyed by touching sample edges no matter how fast it moves. However, for the FM skyrmion, there is a transverse drift, i.e., the skyrmion Hall effect [68-70], which may cause the skyrmion to be destroyed at the sample edge [71, 72]. Various ways have been proposed to overcome or suppress the skyrmion Hall effect, such as adopting an antiferromagnetically coupled bilayer geometry [45], using a curbed racetrack [72], or applying high PMA in the racetrack edge [71]. Second, the speed of an AFM skyrmion is larger than that of a FM skyrmion under the same driving force.

These two advantages of the AFM skyrmion are due to the cancellation of the Magnus force. Since the magnetizations in two sublattices are opposite, an AFM skyrmion can be seen as the combination of two skyrmions with the opposite ferromagnetic topological number $Q$, [1, 6, 42, 43, 73]

$$Q = -\frac{1}{4\pi} \int dxdy [\bm{m} \cdot (\partial_x \bm{m} \times \partial_y \bm{m})], \quad (15)$$

which equals ± 1 for an isolated skyrmion. The opposite Magnus forces ($4\pi Q \bm{e}_z \times \bm{v}$) acted on the skyrmion in each sublattice are counteracted perfectly, thus there is no skyrmion Hall effect, as reported in Refs. [42, 43]. In addition, based on Thiele equation, the velocity of an AFM skyrmion is much larger than that of a FM skyrmion. It should be mentioned that the steady velocities of AFM skyrmions and FM skyrmions induced by a magnetic anisotropy gradient are given by the same formula (see Ref. [63] for details), Eq. (6), if the Magnus force acting on the FM skyrmion is balanced by the forces of the surrounding environment. However, the FM skyrmion will be destroyed by touching the edge when it moves too fast due to large Magnus force. Our results can be extended to the antiferromagnetically exchange coupled bilayer system. As reported in Ref. [45], the Magnus forces acting on the skyrmions in the top and bottom FM layers are exactly cancelled when the interlayer AFM exchange coupling is strong enough. Therefore, for this case of strong interlayer AFM exchange coupling, the steady velocity of the AFM bilayer skyrmion driven by a magnetic anisotropy gradient can also be calculated by Eq. (6).

In summary, we demonstrate that the AFM skyrmion can be driven by a magnetic anisotropy gradient. This motion has been investigated analytically and numerically. The results show that its velocity is proportional to $1/\alpha$ and $dK/dx$ and can reach ~500 m/s. We also give an expression which reveals the effect of skyrmion size on the motion of an AFM skyrmion, it shows that an AFM skyrmion with larger size obtains higher velocity. The motion of AFM and FM skyrmions driven by an anisotropy gradient are also compared. It is found that the velocity of an AFM skyrmion is much larger than that of a FM skyrmion without showing skyrmion Hall effect. Our results provide a promising way to manipulate AFM textures for future spintronic applications.

## ACKNOWLEDGMENTS

G.Z. acknowledges the support by the National Natural Science Foundation of China (Grant Nos. 51771127, 51571126 and 51772004) of China, the Scientific Research Fund of Sichuan Provincial Education Department (Grant Nos. 18TD0010 and 16CZ0006). X.Z. was supported by JSPS RONPAKU (Dissertation Ph.D.) Program. M.E. acknowledges the support by the Grants-in-Aid for Scientific Research from JSPS KAKENHI (Grant Nos. JP18H03676, JP17K05490 and JP15H05854), and also the support by CREST, JST (Grant No. JPMJCR16F1). O.A.T. acknowledges support by the Grants-in-Aid for Scientific Research (Grant Nos. 17K05511 and 17H05173) from MEXT, Japan and by JSPS and RFBR under the Japan-Russian Research Cooperative Program. X.L. acknowledges the support by the Grants-in-Aid for Scientific Research from JSPS KAKENHI (Grant Nos. 17K19074, 26600041 and 22360122). Y.Z. acknowledges the support by the President's Fund of CUHKSZ, the National Natural Science Foundation of China (Grant No. 11574137), and Shenzhen Fundamental Research Fund (Grant Nos. JCYJ20160331164412545 and JCYJ20170410171958839).

arXiv:1808.08664 [cond-mat.mes-hall][56] M. C. T. Ma, Y. Xie, H. Sheng, S. J. Poon, and A. Ghosh, arXiv preprint arXiv:1806.06334 (2018).

[57] R. Zarzuela and Y. Tserkovnyak, Phys. Rev. B **95**, 180402(R) (2017).

[58] A. N. Bogdanov, U. K. Rößler, M. Wolf, and K. H. Müller, Phys. Rev. B **66**, 214410 (2002).

[59] T. Maruyama, Y. Shiota, T. Nozaki, K. Ohta, N. Toda, M. Mizuguchi, A. A. Tulapurkar, T. Shinjo, M. Shiraishi, S. Mizukami, Y. Ando, and Y. Suzuki, Nat. Nanotech. **4**, 158 (2009).

[60] S. Yoichi, M. Takuto, N. Takayuki, S. Teruya, S. Masashi, and S. Yoshishige, Appl. Phys. Express **2**, 063001 (2009).

[61] S. Rohart and A. Thiaville, Phys. Rev. B **88**, 184422 (2013).

[62] T. L. Gilbert, IEEE Trans. Magn. **40**, 3443 (2004).

[63] See Supplemental Material at http://link.aps.org/supplemental/ for details on the analytical velocity of an AFM skyrmion and a FM skyrmion driven by a PMA gradient, micromagnetic simulations for the PMA gradient-induced motion a FM skyrmion and current-induced motion of an AFM skyrmion.

[64] S. S. Parkin, M. Hayashi, and L. Thomas, Science **320**, 190 (2008).

[65] J. Wang, J. Xia, X. Zhang, G. P. Zhao, L. Ye, J. Wu, Y. Xu, W. Zhao, Z. Zou, and Y. Zhou, J. Phys. D: Appl. Phys. **51**, 205002 (2018).

[66] A. A. Thiele, Phys. Rev. Lett. **30**, 230 (1973).

[67] S.-Z. Lin, Phy. Rev. B **96**, 014407 (2017).

[68] K. Litzius, I. Lemesh, B. Krüger, P. Bassirian, L. Caretta, K. Richter, F. Büttner, K. Sato, O. A. Tretiakov, J. Förster, R. M. Reeve, M. Weigand, I. Bykova, H. Stoll, G. Schütz, G. S. D. Beach, and M. Kläui, Nat. Phys. **13**, 170 (2017).

[69] W. Jiang, X. Zhang, G. Yu, W. Zhang, X. Wang, M. Benjamin Jungfleisch, J. E. Pearson, X. Cheng, O. Heinonen, K. L. Wang, Y. Zhou, A. Hoffmann, and S. G. E. te Velthuis, Nat. Phys. **13**, 162 (2017).

[70] I. A. Ado, O. A. Tretiakov, and M. Titov, Phys. Rev. B **95**, 094401 (2017).

[71] P. Lai, G. P. Zhao, H. Tang, N. Ran, S. Q. Wu, J. Xia, X. Zhang, and Y. Zhou, Sci. Rep. **7**, 45330 (2017).

[72] I. Purnama, W. L. Gan, D. W. Wong, and W. S. Lew, Sci. Rep. **5**, 10620 (2015).

[73] O. A. Tretiakov and O. Tchernyshyov, Phys. Rev. B **75**, 012408 (2007).
Page **16** of **19**

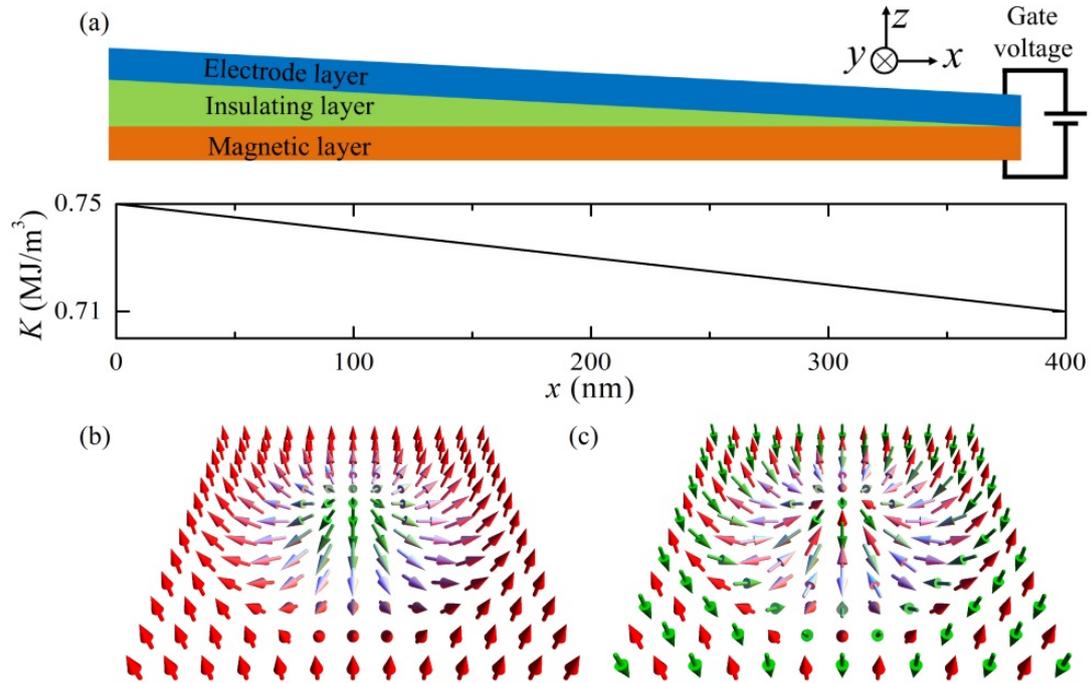

FIG. 1. (a) A sketch of the voltage-controlled magnetic anisotropy device, where the voltage-controlled magnetic anisotropy $K$ linearly decreases with the increase of the spacial coordinate $x$ for $K_0 = 0.75$ MJ/m$^3$ and $dK/dx = 100$ GJ/m$^4$. The spin textures of (b) a ferromagnetic skyrmion and (c) an antiferromagnetic skyrmion. The magnetic parameters are adopted from Ref. [14], that is, $M_S = 580$ kA/m, $K_0 = 0.75$ MJ/m$^3$ ~ $0.85$ MJ/m$^3$, $A = 15$ pJ/m, $D = 4$ mJ/m$^2$, $\alpha = 0.002$ ~ $0.2$, $\gamma = 2.211 \times 10^5$ m/(A s). In the simulation, we employed a square lattice model. The model size is $400 \times 100 \times 0.6$ nm$^3$ and mesh size is $1 \times 1 \times 0.6$ nm$^3$.

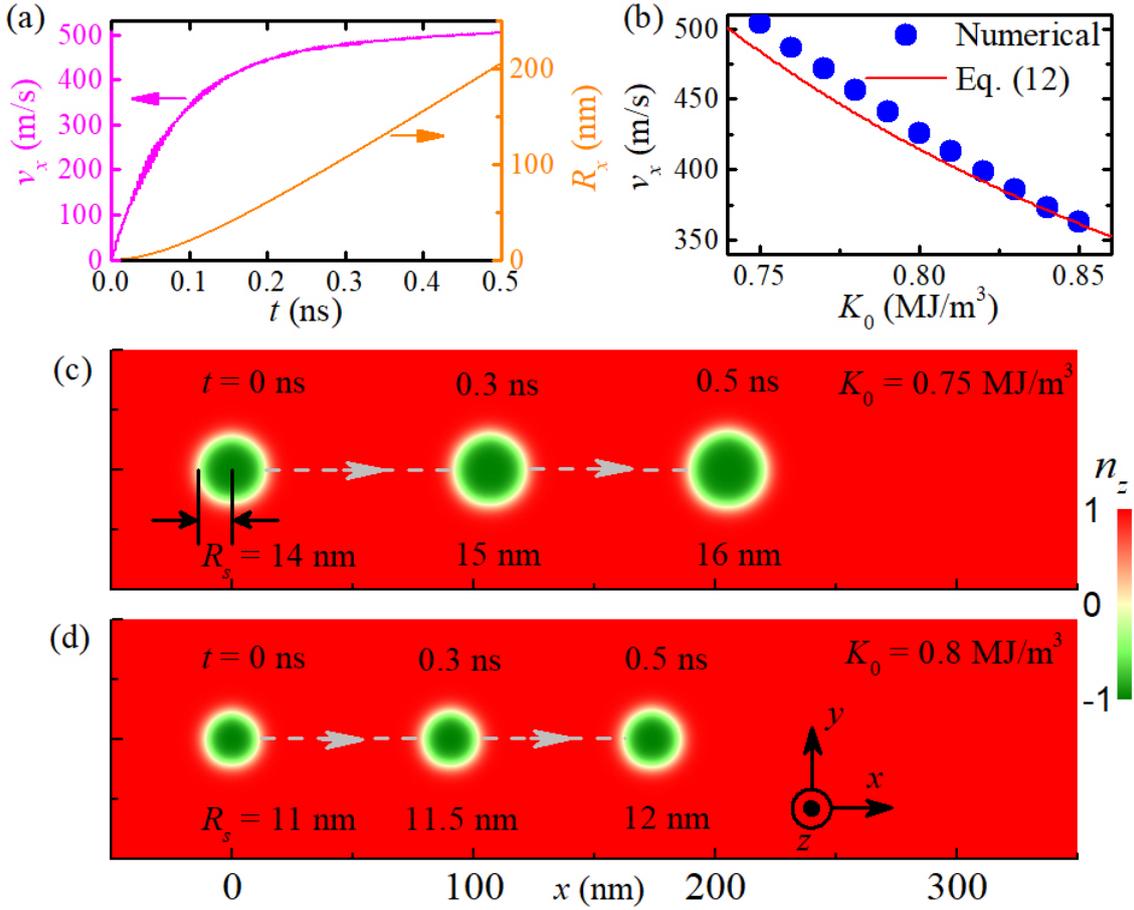

FIG. 2. (a) The evolution of velocity $v_x$ and position $R_x$ for an AFM skyrmion induced by a magnetic anisotropy gradient for $K_0 = 0.75$ MJ/m$^3$. (b) The influence of magnetic anisotropy $K_0$ on velocity $v_x$, where symbols stand for the numerical velocity at $t = 0.5$ ns, whereas the line is calculated using Eq. (12). (c)-(d) The top-views of the AFM skyrmion motion with $K_0$ = 0.75 MJ/m$^3$ and 0.8 MJ/m$^3$, respectively. In the simulation, anisotropy gradient $dK/dx$ = 100 GJ/m$^4$ is used as the driving source, $A_h l^2 = 10$ MJ/m$^3$ is assumed, and the damping constant $\alpha$ (= $G_1 l = G_2/l$) = 0.002.

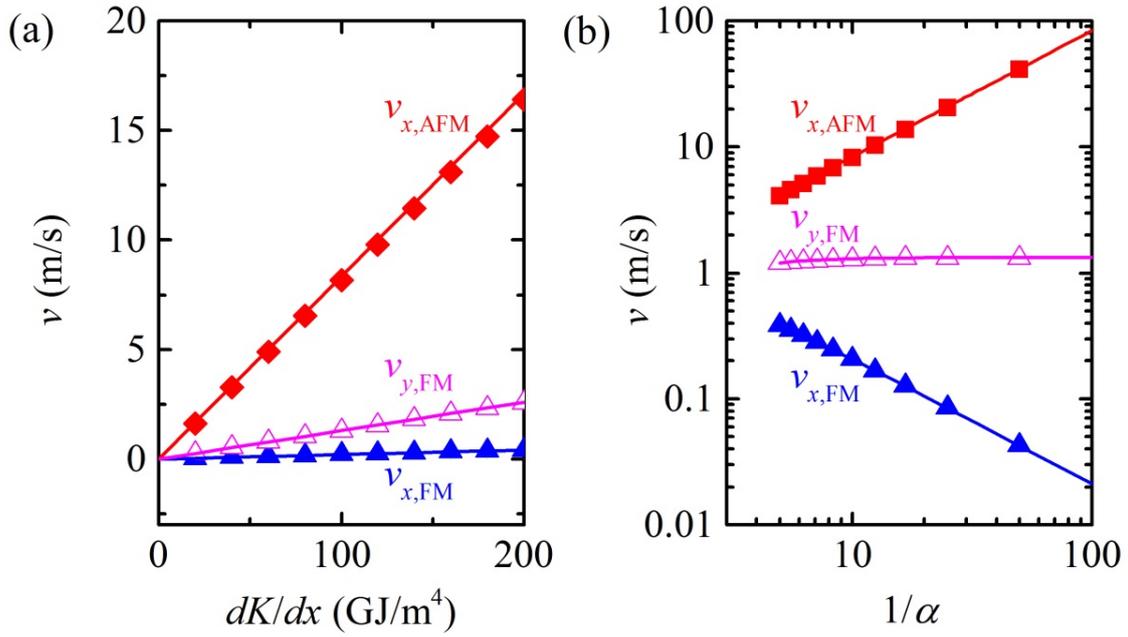

FIG. 3. The numerical (symbols) and analytical (lines) velocities of an AFM skyrmion and a FM skyrmion. (a) The velocities as functions of anisotropy gradient $dK/dx$ for $K_0 = 0.8$ MJ/m$^3$ and $\alpha = 0.1$. (b) The velocities as functions of $1/\alpha$ for $K_0 = 0.8$ MJ/m$^3$ and $dK/dx = 100$ GJ/m$^4$. Here, the numerical velocities are obtained at $t = 0.2$ ns, whereas the analytical results are based on Eqs. (6) and (14).